\documentclass[prl,twocolumn]{revtex4}%
\usepackage{amsmath}
\usepackage{amssymb}
\usepackage{amsfonts}
\usepackage{graphicx}%
\begin{document}
\title{Photoexcited electron dynamics in Kondo insulators and heavy fermions}
\author{Jure Demsar$^{1}$, Verner K. Thorsm\o lle$^{2}$, John L. Sarrao$^{2}$,\ and
Antoinette J. Taylor$^{2}$}
\affiliation{$^{1}$Jozef Stefan Institute, Jamova 39, SI-1000, Ljubljana, Slovenia}
\affiliation{$^{2}$Los Alamos National Laboratory, Los Alamos, New Mexico 87545, USA}

\begin{abstract}
We have studied the photoexcited carrier relaxation dynamics in
the Kondo insulator SmB$_{6}$ and the heavy fermion metal
YbAgCu$_{4}$ as a function of temperature and excitation level.
The dynamic response is found to be both strongly temperature
dependent and nonlinear. The data are analyzed with a
Rothwarf-Taylor bottleneck model, where the dynamics are governed
by the presence of a narrow gap in the density of states near the
Fermi level. The remarkable agreement with the model suggests that
carrier relaxation in a broad class of heavy electron systems
(both metals and insulators) is governed by the presence of a
(weakly temperature dependent) hybridization gap .

\end{abstract}
\maketitle

Recent experiments on the photoexcited electron dynamics in heavy fermion
metals have shown that the picosecond (ps) relaxation time of the photoinduced
(PI) reflectivity increases by more than two orders of magnitude upon cooling
from 300 to 10 K \cite{DemsarHF,Ahn}. In contrast, the dynamics of
their\ non-magnetic analogues were similar to conventional metals like Au and
Ag \cite{Groenevald}, and could be described by the so-called two-temperature
model (TTM) \cite{Groenevald,Hase}, where the ps recovery is governed by
electron-phonon (e-ph) thermalization. The relaxation times in magnetic and
non-magnetic compounds were quite similar at high temperatures (T) suggesting
the TTM as a starting point for understanding the relaxation processes also in
heavy fermions. The TTM analysis of the dynamics showed that the relaxation
process in heavy fermions can be accounted for using a simple e-ph
thermalization, assuming that there exists a mechanism for e-ph scattering
suppression when both the initial and final electronic states lie within the
peak in the density of states (DOS) at the Fermi level (E$_{f}$)
\cite{DemsarHF,Ahn}. It was argued that the small Fermi velocity compared to
the sound velocity may be the origin of this suppression, because, in this
case, the energy and momentum conservation law suppresses e-ph thermalization
\cite{Ahn}.

In the previous work, there were some observations that were not described by
the TTM model, such as the anomalous rise-time dynamics observed in
YbAgCu$_{4}$ at low T \cite{DemsarHF}. Importantly, similar T-dependent
rise-rime dynamics have also been observed in superconductors
\cite{DemsarMgB2}. Furthermore, the T-dependence of the relaxation rate
observed in YbAgCu$_{4}$ also closely resembles the data on superconductors,
both conventional \cite{DemsarMgB2} and cuprate \cite{Schneider,mercury}. The
very similar behavior of the PI carrier dynamics in superconductors and heavy
electron systems suggests that the physics governing the relaxation dynamics
in heavy electron compounds\ may need to be reconsidered. Indeed, if
hybridization of the local \textsf{f}-moments with conduction electrons leads
to the opening of a well established (indirect) hybridization gap (E$_{g}$)
near E$_{f}$ \cite{Riseborough,Reviews}, similar relaxation bottleneck is
expected as observed in superconductors \cite{KabanovRT}. The analogy with
superconductors is straightforward for the case of Kondo insulators, where
E$_{f}$ lies within the gap \cite{Riseborough}. However, the same bottleneck
physics \cite{KabanovRT} is expected to be effective even in the case of
metallic heavy fermions, if E$_{f}$ lies close to the hybridization gap edge
(if the distance between the gap edge and E$_{f}$ $\ll E_{g}$).

Relaxation phenomena for nonequilibrium superconductors have revealed some of
the most intriguing problems in condensed matter physics since the 1960's
\cite{NonSup}. The presence of the superconducting gap in the single particle
excitation spectrum presents a relaxation bottleneck. The minimal model that
describes the relaxation of the (photo)excited superconductor was formulated
in 1967 by Rothwarf and Taylor \cite{roth}. Pointing out that the phonon
channel should be considered when discussing relaxation processes, they
described the relaxation dynamics by two coupled nonlinear-differential
equations \cite{roth}. While in the low perturbation limit the equations can
be linearized \cite{Kaplan}, it was shown only recently that approximate
analytical solutions can be obtained for all limiting cases \cite{KabanovRT}.
Analytical solutions enable comparison of the experimental data with the
model, and have revealed that the Rothwarf-Taylor (RT) model can account for
both the rise-time dynamics \cite{DemsarMgB2} as well as the superconducting
state recovery as a function of excitation fluence (\emph{F}) and T
\cite{KabanovRT}.

In this Letter, we present the first detailed study of the carrier relaxation
dynamics in heavy electron systems as a function of excitation level.
Utilizing a low repetition rate optical parametric amplifier we were able to
measure the \emph{F}-dependence of the reflectivity transients in the Kondo
insulator SmB$_{6}$ and heavy fermion YbAgCu$_{4}$ over more than three orders
of magnitude in \emph{F}. Both the transient amplitude and the relaxation rate
were found to be strongly \emph{F-}dependent. Moreover, both observations are
found to be consistent with the RT model implying that the relaxation
phenomena in Kondo insulators, as well as in heavy fermion metals, are
governed by a phonon bottleneck mechanism arising from the presence of the
hybridization gap in the DOS at E$_{f}$. The temperature dependence of the
transient amplitude and relaxation rate further supports this scenario, and
the extracted values of the hybridization gap are in good agreement with the
published data. The data suggest that the hybridization gap is, to a first
approximation, T-independent consistent with the periodic Anderson lattice
model in the $U=0$ limit \cite{Reviews,Joyce}.

In order to study the \emph{F-}dependence of the dynamical response at low T
while minimizing the continuous heating of the probed spot due to laser
excitation, we used an amplified Ti:Al$_{2}$O$_{3}$ laser system and an
optical parametric amplifier operating at 250 kHz producing sub 100 fs pulses.
The experiments were performed using a standard pump-probe set-up. The samples
were excited at 3.0~eV with \emph{F} ranging from 0.5 to 600 $\mu$J/cm$^{2}$,
while the PI changes in reflectivity were measured at a photon energy of 1.67
eV. The lowest \emph{F} used in this study was only about a factor of 5 higher
than \emph{F} used in an earlier study \cite{DemsarHF}, while the maximum
\emph{F} is more than 1000 times higher. Moreover, since the repetition rate
of the amplified system is 3 orders of magnitude lower than in the high
repetition rate system \cite{DemsarHF}, the effect of sample heating is
minimized.%
%TCIMACRO{\FRAME{fhFU}{8.5536cm}{5.3246cm}{0pt}{\Qcb{(a) The \emph{F}
%dependence of the PI reflectivity taken in YbAgCu$_{4}$ at 5 K. The data are
%normalized to \emph{F} to emphasize the sub-linear \emph{F}-dependence of the
%amplitude \textsf{A}. (b) The T-dependence of the PI reflectivity taken at
%$\emph{F}=8.4$ $\mu$J/cm$^{2}$. The inset to (a) illustrates the way \textsf{A
%}and $\tau$ are extracted from the raw data; the data set taken at 5 K and
%\emph{F }$=0.28$ $\mu$J/cm$^{2}$ serves as an example.}}{}{fig1.eps}%
%width 8.5536cm;  height 5.3246cm;  depth 0pt;  original-width 4.1079in;
%original-height 2.9222in;  cropleft "0.0363";  croptop "0.8804";
%cropright "0.9319";  cropbottom "0.0581";
%filename 'Fig1.eps';file-properties "XNPEU";}}}%
%BeginExpansion
\begin{figure}
[h]
\begin{center}
\includegraphics[
trim=0.149117in 0.169780in 0.279748in 0.349495in,
height=5.3246cm,
width=8.5536cm
]%
{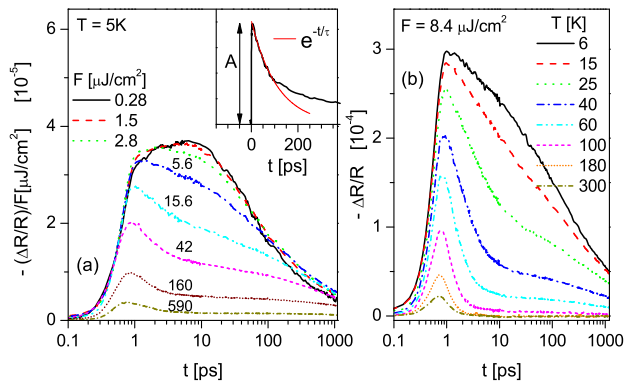}%
\caption{(a) The \emph{F} dependence of the PI reflectivity taken in
YbAgCu$_{4}$ at 5 K. The data are normalized to \emph{F} to emphasize the
sub-linear \emph{F}-dependence of the amplitude \textsf{A}. (b) The
T-dependence of the PI reflectivity taken at $\emph{F}=8.4$ $\mu$J/cm$^{2}$.
The inset to (a) illustrates the way \textsf{A }and $\tau$ are extracted from
the raw data; the data set taken at 5 K and \emph{F }$=0.28$ $\mu$J/cm$^{2}$
serves as an example.}%
\end{center}
\end{figure}
%EndExpansion

Figure 1(a) presents the PI reflectivity in YbAgCu$_{4}$ measured at 5 K while
\emph{F} is varied from 0.28 to 580 $\mu$J/cm$^{2}$. All the reflectivity
traces in panel (a) are normalized to \emph{F},\emph{ }i.e. $\Delta$R$/$R is
divided by F\ in units of $\mu$J/cm$^{2}$. For a linear response, all the
traces normalized to \emph{F }would lie on top of each other; this is clearly
not the case. The inset shows the data taken at 0.28 $\mu$J/cm$^{2}$ on a
linear scale. The recovery dynamics is found to be non-exponential, and the
relaxation non-universal, which is to be expected in the non-linear regime.
Therefore, when performing a quantitative analysis of the data, we analyze the
amplitude of the transients, \textsf{A,} and the initial relaxation rate,
$\tau^{-1}$, extracted by fitting the early relaxation dynamics with
exponential decay (see inset to panel (a)).

At the lowest \emph{F}, the data are consistent with the low T data obtained
in an earlier study \cite{DemsarHF}. The initial femtosecond rise-time
($\sim200$ fs, resolution limited) is followed by a further increase of the
signal on a ps timescale. The recovery dynamics proceeded on a 100 ps
timescale. As \emph{F} is increased, the PI transient is changed dramatically.
The second-stage risetime becomes faster and above \emph{F }$\approx4$ $\mu
$J/cm$^{2}$ the rising edge is resolution limited. The decay rate is also
weakly \emph{F}-dependent up to $\approx4$ $\mu$J/cm$^{2}$ and increases by
almost 2 orders of magnitude as \emph{F} is further increased to the 100 $\mu
$J/cm$^{2}$ range. Moreover, the amplitude \textsf{A} shows linear
\emph{F-}dependence up to \emph{F }$\approx4$ $\mu$J/cm$^{2}$, while at higher
fluences \textsf{A }$\approx\sqrt{F}$. Panel (b) shows the T-dependence of the
reflectivity transient taken at constant $\emph{F}$. The behavior is similar
to earlier low \emph{F} data \cite{DemsarHF}, revealing two orders of
magnitude increase in the relaxation rate upon warming to 300 K, while
\textsf{A} gradually decreases.

Figure 2 presents \emph{F}- and T-dependence of the PI reflectivity in the
Kondo insulator SmB$_{6}$ \cite{Riseborough}. The behavior is very similar to
YbAgCu$_{4}$. The main difference is that in SmB$_{6}$ no anomalous two-stage
risetime is observed. The fluence at which \textsf{A} departs from
$\mathsf{A}\varpropto\emph{F}$ is\ near\emph{ }$12$ $\mu$J/cm$^{2}$.%
%TCIMACRO{\FRAME{fhFU}{8.6525cm}{5.5113cm}{0pt}{\Qcb{(a) The \emph{F}%
%-dependence of the PI reflectivity in the Kondo insulator SmB$_{6}$ at 5 K.
%The data have been normalized to \emph{F} in order to emphasize the sub-linear
%\emph{F}-dependence of the amplitude. (b) The T-dependence of the PI
%reflectivity taken at $\emph{F}=8.4$ $\mu$J/cm$^{2}.$}}{}{fig2.eps}%
%width 8.6525cm;  height 5.5113cm;  depth 0pt;  original-width 4.1079in;
%original-height 2.9222in;  cropleft "0.0315";  croptop "0.8906";
%cropright "0.9441";  cropbottom "0.0716";
%filename 'Fig2.eps';file-properties "XNPEU";}}}%
%BeginExpansion
\begin{figure}
[h]
\begin{center}
\includegraphics[
trim=0.129399in 0.209230in 0.229632in 0.319689in,
height=5.5113cm,
width=8.6525cm
]%
{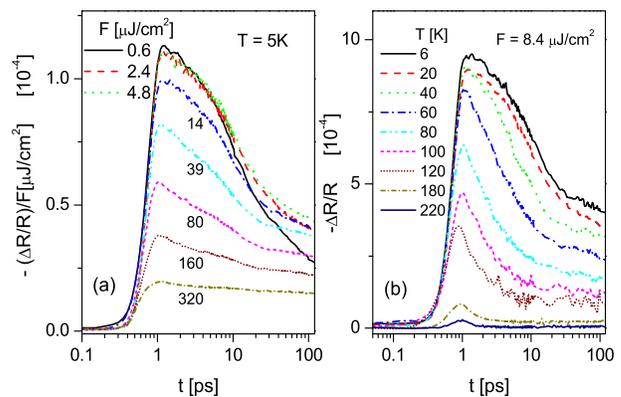}%
\caption{(a) The \emph{F}-dependence of the PI reflectivity in the Kondo
insulator SmB$_{6}$ at 5 K. The data have been normalized to \emph{F} in order
to emphasize the sub-linear \emph{F}-dependence of the amplitude. (b) The
T-dependence of the PI reflectivity taken at $\emph{F}=8.4$ $\mu$J/cm$^{2}.$}%
\end{center}
\end{figure}
%EndExpansion

Before analyzing the \emph{F}- and T-dependence of the PI reflectivity
transients, we briefly review the RT model and its solutions, using language
appropriate for a photoexcited narrow gap semiconductor. Following
photoexcitation, high energy electrons initially release their energy via e-e
and e-ph scattering. After this process, which usually proceeds on a sub-ps
timescale \cite{Groenevald}, the system is characterized by excess densities
of electron-hole pairs (EHP) and high frequency phonons (HFP). When an EHP
with an energy $\geq E_{g}$ ($E_{g}$ is the gap) recombines a high frequency
phonon ($\omega>E_{g}$) is created. Since HFP can subsequently excite EHP, the
recovery is governed\ by the decay of the HFP population \cite{roth}. The
dynamics of EHP and HFP populations is determined by \cite{roth}
\begin{align}
dn/dt  &  =\eta N-Rn^{2},\nonumber\\
dN/dt  &  =-\eta N/2+Rn^{2}/2-\gamma(N-N_{T}). \label{RTeq}%
\end{align}
Here $n$ and $N$ are the concentrations of EHP and HFPs, respectively, $\eta$
is the probability for EHP creation by HFP absorption, and $R$ the rate of
electron-hole recombination with the creation of a HFP. $N_{T}$ is the
concentration of HFPs in thermal equilibrium, and $\gamma$ their decay rate
(governed\ either by anharmonic decay or by diffusion out of the excitation volume).

The thermal equilibrium concentrations of HFPs and EHPs ($n_{T}$)
satisfy the detailed balance equation $Rn_{T}^{2}=\beta N_{T}$.
Depending on the initial conditions ($n_{0}$ and $N_{0}$, which
are concentrations of EHP and HFP after photoexcitation and the
initial e-e and e-ph avalanche process) and the ratio of
$\gamma/\eta$, several different regimes are realized
\cite{KabanovRT}. It follows from the RT analysis that the ps
risetime dynamics is observed at low \emph{F} when photoexcitation
and the initial avalanche processes lead to an excess phonon
population with respect to the detailed balance condition; i.e.
$Rn_{0}^{2}<\beta N_{0}$ \cite{KabanovRT}. This is followed by the
"thermalization" of EHP and HFP distributions leading to
quasi-stationary distributions of $n_{s}$ and $N_{s}$ that satisfy
\cite{DemsarMgB2,KabanovRT}
\begin{equation}
Rn_{s}^{2}=\beta N_{s}\text{ ;\ }n_{s}=\frac{R}{4\eta}(\sqrt{1+\frac{16R}%
{\eta}n_{0}+\frac{8R}{\eta}N_{0}}-1). \label{stationary}%
\end{equation}
Since the PI reflectivity (absorption) is proportional to the PI carrier
density, the resulting PI reflectivity shows an initial fast rise-time ($n(t)$
reaches $n_{0}$) followed by a ps rise where $n(t)$ reaches $n_{s}$. Recovery
proceeds on a much longer timescale and is governed by HFP decay.

There are several important predictions of the model that can be used to
determine whether photoexcited carrier dynamics in heavy electron systems
indeed follow the RT kinetics. First, we should note that the dynamics at low
T should be strongly \emph{F-}dependent. Since the amplitude \textsf{A} is a
measure of the PI e-h density, i.e. \textsf{A }$\varpropto n_{s}-n_{T}$, and
$n_{0}$ and $N_{0}$ are, at low-T, proportional to \emph{F}, it follows from
Eq.(\ref{stationary}) that \textsf{A }$\varpropto\sqrt{1+c\emph{F}}-1$, where
$c$ is a constant. As seen in Fig. 3(a) the \emph{F-}dependence of \textsf{A}
for both SmB$_{6}$ and YbAgCu$_{4}$ is in excellent agreement with the model.

The initial relaxation rate $\tau^{-1}$ is given by \cite{KabanovRT}
\begin{equation}
\tau^{-1}=\frac{2R\gamma(n_{s}+n_{T})}{\eta^{2}(1+2\gamma/\eta)}\text{.}
\label{lowTtau}%
\end{equation}
It suggests that at low \emph{F},\textsf{\ }when $n_{s}\approx n_{T}$, the
relaxation rate should be \emph{F}-independent, while at high fluences,
$\tau^{-1}\varpropto\emph{F}$. As shown in Fig. 3(b) the data for SmB$_{6}$
agree well with the model over the entire range of \emph{F}. In YbAgCu$_{4}$,
on the other hand, at high \emph{F} the experimental $\tau^{-1}$ increases
faster than the theoretical prediction, finally saturating. Saturation is also
observed in \textsf{A}(\emph{F})\textsf{\ }for both compounds. This saturation
can be attributed to the PI smearing of the hybridization gap structure.%
%TCIMACRO{\FRAME{fhFU}{8.7052cm}{5.3576cm}{0pt}{\Qcb{The \emph{F}-dependence of
%(a) the amplitude \textsf{A} and (b) the initial relaxation rate $\tau^{-1}$
%for SmB$_{6}$ and YbAgCu$_{4}$ at 5 K. Solid lines are fits to the data with
%the RT model, $\mathsf{A}\varpropto\sqrt{1+c\emph{F}}-1$ and $\tau
%^{-1}\varpropto(n_{s}+n_{T})$ (see text), while dashed lines in panel (a)
%represent the best linear fit (arrows indicate the departure from
%$\mathsf{A}\varpropto\emph{F}$ dependence).}}{}{fig3.eps}%
%width 8.7052cm;  height 5.3576cm;  depth 0pt;  original-width 3.2776in;
%original-height 2.3134in;  cropleft "0.0426";  croptop "0.8488";
%cropright "0.9513";  cropbottom "0.0604";
%filename 'Fig3.eps';file-properties "XNPEU";}}}%
%BeginExpansion
\begin{figure}
[h]
\begin{center}
\includegraphics[
trim=0.139626in 0.139729in 0.159619in 0.349786in,
height=5.3576cm,
width=8.7052cm
]%
{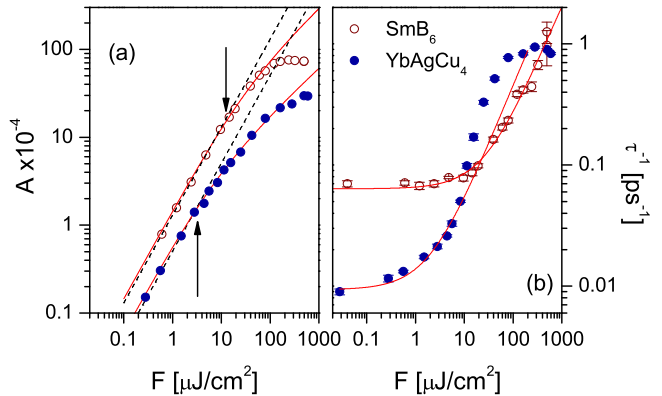}%
\caption{The \emph{F}-dependence of (a) the amplitude \textsf{A} and (b) the
initial relaxation rate $\tau^{-1}$ for SmB$_{6}$ and YbAgCu$_{4}$ at 5 K.
Solid lines are fits to the data with the RT model, $\mathsf{A}\varpropto
\sqrt{1+c\emph{F}}-1$ and $\tau^{-1}\varpropto(n_{s}+n_{T})$ (see text), while
dashed lines in panel (a) represent the best linear fit (arrows indicate the
departure from $\mathsf{A}\varpropto\emph{F}$ dependence).}%
\end{center}
\end{figure}
%EndExpansion

The present \emph{F}-dependence studies clearly support the idea of electron
relaxation dynamics in heavy electron compounds being governed by the presence
of a hybridization gap in the DOS. Moreover, as was shown in Ref.
\cite{KabanovRT}, the T-dependencies (at constant \emph{F}) of both \textsf{A}
and $\tau^{-1}$ are governed by the T-dependence of the number density of
thermally excited EHP's, $n_{T}$. It was shown that $n_{T}\propto
\mathcal{A}^{-1}-1$, where $\mathcal{A}(T)=$ \textsf{A}$(T)/$\textsf{A}%
$(T\longrightarrow0)$. It is easy to demonstrate that the T-dependence of
$\tau^{-1}$ is governed by the T-dependence of $n_{T}$ as well. Assuming
$R,\gamma,$ and $\eta$ to be T-independent, Eq.(\ref{lowTtau}) can be
rewritten as $\tau^{-1}(T)=C[(n_{s}-n_{T})+2n_{T}]$, where $C$ is a
proportionality constant. Furthermore, since \textsf{A }$\varpropto
n_{s}-n_{T}$ and $n_{T}\propto\mathcal{A}^{-1}-1$, it follows that
\begin{equation}
\tau^{-1}(T)=C[D(n_{T}+1)^{-1}+2n_{T}]\text{ \ ,} \label{TauTdep}%
\end{equation}
where $D$ is a constant that depends only on the photoexcitation intensity.
Therefore, for constant \emph{F}, $\tau^{-1}(T)$ is entirely determined by the
T-dependence of $n_{T}$.

In a narrow band semiconductor the T-dependence of $n_{T}$ depends on the
shape of the DOS in the energy range $\varepsilon\approx kT$ around the
chemical potential. Generally, $n_{T}$ is given by:
\begin{equation}
n_{T}\simeq T^{p}\exp(-E_{g}/2T), \label{nTEg}%
\end{equation}
where $p$ is on the order of 1, depending on the exact shape of the DOS near
the gap edge. For a hybridization gap scenario, the shape of the DOS should be
close to that of a BCS superconductor, where $p=1/2$. Neither the exact shape
of the low energy DOS, nor T-dependence of $E_{g}$ is well known in heavy
electron systems, therefore there will be some ambiguity in determining the
precise value of the indirect gap $E_{g}$. However, since the main
T-dependence in Eq. [\ref{nTEg}] comes from the exponential term, a rough
estimate of the size and the T-dependence of $E_{g}$ can be obtained.

In Figure 4(a), we plot the T-dependence of $n_{T}$ for the two compounds. In
both data sets $n_{T}$ changes by almost 3 orders of magnitude between $20$
and $200$ K. Moreover, the T-dependence of $n_{T}$ is far from linear
(expected in the TTM scenario \cite{Hase}) over the entire T-range.
Furthermore, the absence of any discontinuity or change of slope in $n_{T}$ at
high-T suggests that the hybridization gap structure persists to high T, as
observed, for example, in the photoemission spectroscopy of Kondo insulators
\cite{Joyce}. The T-dependence of $\tau^{-1}$ shown in Fig. 4(b) further
supports this conclusion: in SmB$_{6}$, $\tau^{-1}$ increases all the way to
the highest temperatures measured, while in YbAgCu$_{4}$, $\tau^{-1}$
saturates only above $\approx160$ K.%
%TCIMACRO{\FRAME{fhFU}{8.7052cm}{5.0215cm}{0pt}{\Qcb{The T-dependence of (a)
%the density of thermally excited electrons $n_{T}$ $(\propto\mathcal{A}%
%^{-1}-1)$, where $\mathcal{A}$(T) is plotted in the inset, and (b) the initial
%relaxation rate, $\tau^{-1}$. The data were taken at $\emph{F}=8.4$ $\mu
%$J/cm$^{2}$. The dashed lines are fits to the data by the RT model (see
%text).}}{}{fig4.eps}{\special{ language "Scientific Word";  type "GRAPHIC";
%width 8.7052cm;  height 5.0215cm;  depth 0pt;  original-width 3.8303in;
%original-height 2.5313in;  cropleft "0.0468";  croptop "0.8501";
%cropright "0.9453";  cropbottom "0.0709";
%filename 'Fig4.eps';file-properties "XNPEU";}}}%
%BeginExpansion
\begin{figure}
[h]
\begin{center}
\includegraphics[
trim=0.179258in 0.179469in 0.209517in 0.379442in,
height=5.0215cm,
width=8.7052cm
]%
{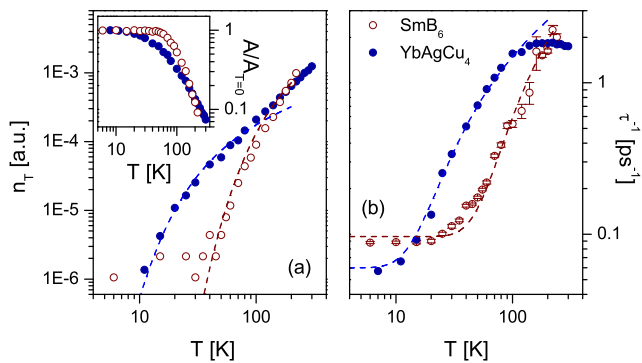}%
\caption{The T-dependence of (a) the density of thermally excited electrons
$n_{T}$ $(\propto\mathcal{A}^{-1}-1)$, where $\mathcal{A}$(T) is plotted in
the inset, and (b) the initial relaxation rate, $\tau^{-1}$. The data were
taken at $\emph{F}=8.4$ $\mu$J/cm$^{2}$. The dashed lines are fits to the data
by the RT model (see text).}%
\end{center}
\end{figure}
%EndExpansion

Based on these qualitative findings, we can apply the RT model to extract the
magnitude of the indirect hybridization gap, $E_{g}$. We consider the
hybridization gap to be present at all T, and to a first approximation,\ to be
T-independent. Furthermore, we assume a BCS-like DOS near the gap edge, i.e.
$p=0.5$. Fitting $n_{T}$ with Eq. (\ref{nTEg}),we find, for SmB$_{6}$, an
excellent agreement over the entire T-range with $E_{g}\approx550$ K. In
YbAgCu$_{4}$, on the other hand, the agreement is good up to $\approx120$ K,
with the extracted value of $E_{g}\approx100$ K. At higher T in YbAgCu$_{4}$,
$n_{T}$ starts to increase faster possibly due to a partial suppression of the
gap at high T.

Qualitative agreement between the relaxation rate data and the model described
by Eq.(\ref{lowTtau}) is also straightforward. At the intermediate
temperatures, $n_{s}-n_{T}\ll n_{T}$, and $\tau^{-1}$ is governed by the
T-dependence of $n_{T}$ exhibiting $\exp(-E_{g}/2T)$ behavior. At low enough
T, however, $n_{s}\gg n_{T}$ and the relaxation time saturates, as observed
experimentally. We fit $\tau^{-1}(T)$ data with Eq. (\ref{TauTdep}) and find
remarkable agreement. The extracted values of the gap are $E_{g}\approx85$ K
for YbAgCu$_{4}$ and $E_{g}\approx350$ K for SmB$_{6}$, somewhat lower than
the values extracted from the fit to $n_{T}$. This can be partially attributed
to the fact that the T-dependence of $R,\gamma,$ and $\eta$ was neglected, as
well as to the microscopic details like gap anisotropy. Comparison with
literature shows that for SmB$_{6}$ there is a large spread of published
values for $E_{g}$ \cite{Nyhus}; however, our estimate is in close agreement
with the T-independent pseudogap energy scale of 290 cm$^{-1}$ (450 K) from a
recent Raman scattering study \cite{Nyhus}. The value of $E_{g}\approx100$ K
for YbAgCu$_{4}$ is also in close agreement with recent optical data
\cite{Hancock}. The overall agreement and self-consistency of the data with
the RT model presents a strong argument that relaxation kinetics in heavy
electron systems is indeed governed by the presence of a weakly T-dependent
hybridization gap, as for a periodic Anderson lattice model in the $U=0$ limit
\cite{Reviews,Joyce}.

We have presented studies of photoexcited electron relaxation and
recombination dynamics in the Kondo insulator SmB$_{6}$ and heavy fermion
YbAgCu$_{4}$ as a function of temperature and excitation level. The dynamical
response is described well using the phenomenological RT model, suggesting
that the carrier relaxation and recombination dynamics in heavy electron
systems are governed by the presence of a weakly T-dependent hybridization gap.

We acknowledge discussions with V. Kabanov and S. Trugman. This work was
supported by US-Slovenian collaboration grant BI-US/05-06/023 and US DOE LDRD program.

\end{document}